\documentclass{article}

 \usepackage[dandb]{neurips_2026}

\usepackage{url}
\usepackage[utf8]{inputenc} 
\usepackage[T1]{fontenc}    
\usepackage{hyperref}       
\usepackage{url}            
\usepackage{booktabs}       
\usepackage{amsfonts}       
\usepackage{nicefrac}       
\usepackage{microtype}      
\usepackage{xcolor}         
\usepackage{graphicx}
\usepackage{booktabs}
\usepackage{amssymb}
\usepackage{pifont}
\usepackage{natbib}
\setcitestyle{numbers}
\setcitestyle{square}
\usepackage{titlesec}
\setcitestyle{comma} 
\usepackage{float}
\usepackage{amsmath}
\newcommand{\cmark}{\ding{51}} 
\newcommand{\xmark}{\ding{55}} 
\usepackage{enumitem}
\setlist[itemize]{leftmargin=*}
\title{SpecX: A Large-Scale Benchmark for Multi-Modal Spectroscopy and Cross-Paradigm Evaluation}
\usepackage{booktabs}
\usepackage{multirow}
\usepackage{makecell}

\usepackage{graphicx}
\usepackage{siunitx} 

\author{
  \begin{tabular}[t]{c}
    Chengrui Xiang \enspace
    Tengfei Ma$^{\dagger}$ \enspace
    Yujie Chen \\
    Tong Wang$^{\dagger}$ \enspace
    Haowen Chen$^{\dagger}$ \enspace
    Xiangxiang Zeng \\[0.5em]
    College of Computer Science and Technology, Hunan University \\
    Changsha, China \\
    \texttt{\{mmtd, tfma, xzeng, wangtong\}@hnu.edu.cn}
  \end{tabular}
  \thanks{$^{\dagger}$Corresponding author}
}

\begin{document}

\maketitle

\begin{abstract}
Existing spectral benchmarks are limited in scale, modality alignment, and evaluation scope, and typically focus on either specialized models or multimodal language models (MLLMs). We introduce SpecX, a large-scale benchmark for multi-modal spectroscopy with cross-paradigm evaluation. SpecX contains 1.7M molecules with diverse spectral modalities, including NMR (¹H, ¹³C, HSQC), IR, MS,UV,Raman and FL, and is organized into three tiers: a large-scale dataset for pretraining, an aligned multi-spectral subset for benchmarking, and a high-quality experimental subset for evaluation. SpecX supports a range of tasks such as molecular elucidation, spectrum simulation, and spectral understanding, and enables unified evaluation across both specialized spectral models and MLLMs. Experiments show that specialized models excel at signal-level modeling, while MLLMs exhibit strengths in high-level reasoning but lack precise spectral grounding. SpecX establishes a unified benchmark for spectral intelligence and highlights the need for spectrum-native foundation models.
\end{abstract}

\section{Introduction}
\label{sec:introduction}

Spectroscopic techniques are fundamental tools for determining molecular 
structure and properties. Different modalities---including Nuclear Magnetic 
Resonance (NMR), Infrared (IR), Mass Spectrometry (MS), Raman, 
Ultraviolet-Visible (UV), and Fluorescence (FL)---each reveal complementary 
aspects of molecular architecture. Just as an experienced chemist integrates 
multimodal spectral information to identify an unknown compound, automated 
structure elucidation (ASE) requires the same integrative reasoning. However, 
achieving robust ASE across diverse chemical spaces remains a significant 
challenge. Recent advances in AI have begun to transform computational 
spectroscopy. Notably, the Multimodal Spectroscopic 
dataset~\cite{alberts2024unraveling} introduced simulated spectra across five 
modalities for approximately 790k molecules, supporting structure elucidation, 
spectrum simulation, and functional group prediction. Despite this progress, 
three critical limitations remain.

First, insufficient scale and modality coverage. Existing datasets remain 
below one million molecules and omit several practically important modalities. 
UV-Vis, fluorescence, and Raman spectroscopy are widely used in chemical 
analysis but largely absent from large-scale benchmarks. Scale also constrains 
the training of spectral foundation models with robust generalization.

Second, lack of experimentally grounded evaluation. Existing large-scale 
datasets rely entirely on simulated spectra, which systematically differ from 
experimental ones due to solvent effects, instrument response, and temperature 
broadening. No existing benchmark provides a systematically aligned experimental subset. SpecX introduces this tier to provide a foundation for future research investigating the simulation-to-reality gap.

Third, fragmented evaluation across model paradigms. As shown in 
Table~\ref{specx}, existing benchmarks fall into two categories: specialized 
model benchmarks~\cite{alberts2024unraveling} (e.g., Multimodal Spec, 
MassSpecGym, NMRNet) focus on signal-level precision but omit MLLMs, while 
MLLM-oriented benchmarks~\cite{guo2024can} (e.g., SpectrumBench, MolPuzzle) 
evaluate high-level reasoning but lack signal-level protocols. A unified 
benchmark enabling side-by-side comparison remains absent.

\begin{figure}[H]
    \centering
    \includegraphics[scale=0.4]{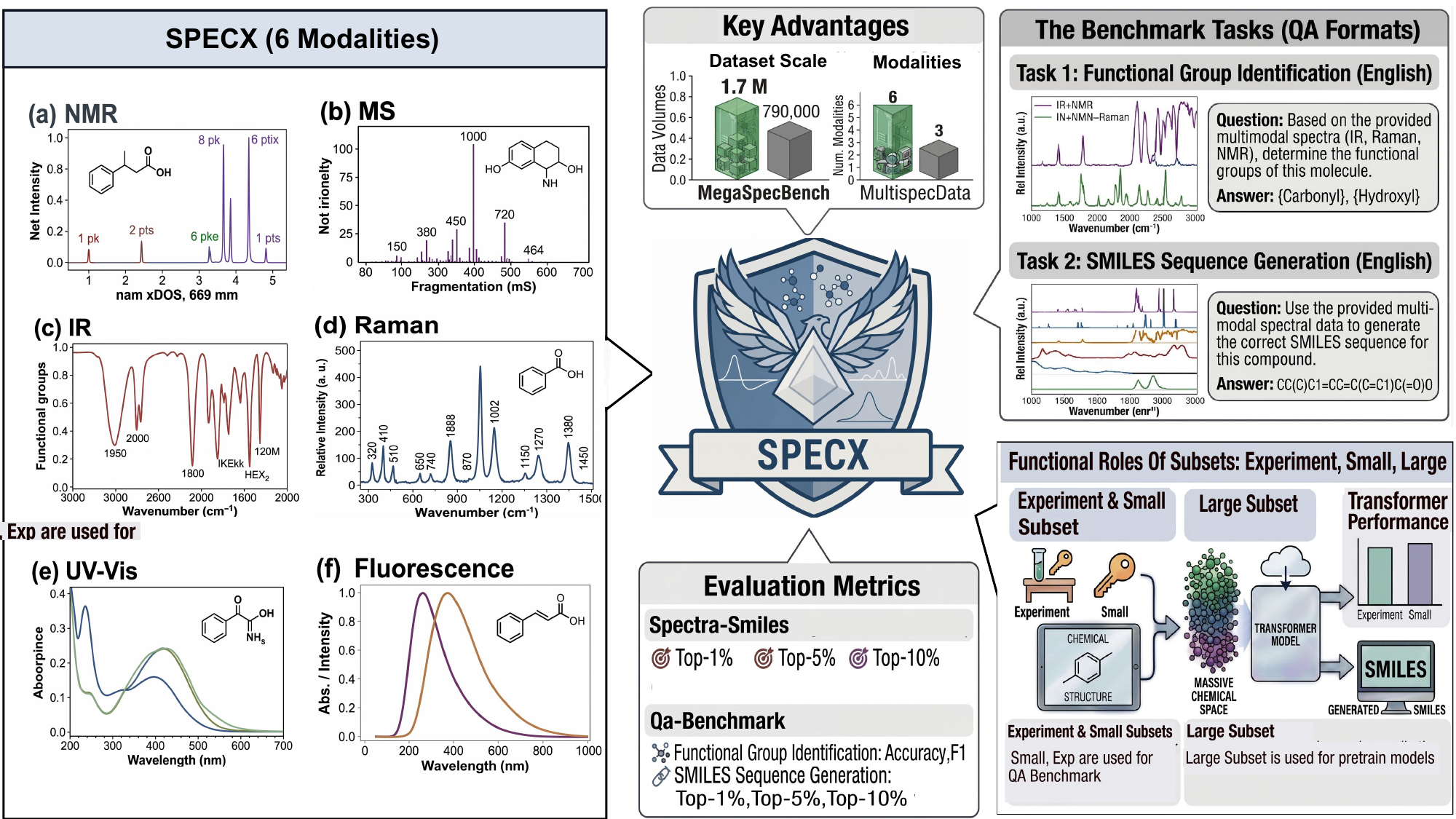}
    \caption{Overview of the SpecX framework.}
    \label{fig:specx_overview}
\end{figure}

To address these gaps, we introduce \textbf{SpecX}, a large-scale multimodal 
spectroscopic benchmark for cross-paradigm evaluation. SpecX contains 1.7 
million molecules spanning eight modalities: $^1$H-NMR, $^{13}$C-NMR, 
HSQC-NMR, IR, MS, UV, FL, and Raman, organized into three tiers: a 
large-scale simulated dataset (Large subset) for pretraining, a 
modality-aligned multispectral subset (Small subset) for multimodal QA, and a 
high-quality experimental subset (Exp subset) for real-world generalization. 
Tasks~(1)--(3) are evaluated on the Large subset; Task~(4) on both the Small 
and Exp subsets. All experiments apply random and scaffold splits to assess 
interpolation and generalization across chemical space.Fluorescence spectra are included in the full dataset but are not part of the current Large/Small
benchmark splits (see Table\ref{tab:subsets}).
For presentation, we also refer to six modality groups by merging the three NMR sub-modalities
($^1$H, $^{13}$C, and HSQC) into a single NMR category (Figure\ref{fig:specx_overview}).

SpecX supports four evaluation tasks: structure elucidation, spectrum 
simulation, functional group QA, and Spectrum-SMILES cross-modal QA, 
integrated at larger scale and broader modality coverage than prior work. 
Experiments reveal complementary strengths: specialized models excel at 
signal-level modeling, while MLLMs demonstrate strong high-level reasoning but 
lack precise spectral grounding. SpecX establishes a unified benchmark for 
spectral intelligence and highlights the need for spectrum-native foundation 
models.

\begin{table}[!ht]
    \centering
    \resizebox{\textwidth}{!}{
    \begin{tabular}{llllll}
    \toprule
        Benchmark & Scale & Multi-Spec & Experimental & Tasks (Elucidation/Simulation) & Model Coverage \\ \midrule
        NovoBench & small & \xmark & \xmark & limited & MLLM \\ 
        MolPuzzle & small & \xmark & \xmark & Elucidation & MLLM \\ 
        Multimodal Spec & medium & \cmark (partial) & partial & Elucidation + Simulation & ML \\ 
        MassSpecGym & medium & \xmark & \cmark & Elucidation + Simulation & ML \\ 
        NMRNet & medium & \xmark & \cmark & Elucidation & ML \\ 
        SpectrumBench & medium & \cmark & \xmark & All & MLLM \\ 
        SpecX (Ours) & \textbf{1.7M} & \cmark (aligned) & \cmark & All & (MLLM + ML) \\ \bottomrule
    \end{tabular}
    }
    \caption{Comparison with existing spectral benchmarks. SpecX provides 
    large-scale, aligned multi-spectral data with experimental validation, and 
    uniquely supports unified evaluation across both specialized spectral models 
    and multimodal language models.}
    \label{specx}
\end{table}
\section{Related Work}
\textbf{Spectral Datasets and Benchmarks:} Spectroscopic datasets form the 
foundation for learning-based molecular analysis. Early single-modality datasets, 
such as nmrshiftdb2 and the NIST IR database, support tasks including peak 
prediction and functional group identification, but are limited in modality 
coverage and do not support cross-modal reasoning. The Multimodal Spectroscopic 
dataset advances this by providing simulated spectra across \textsuperscript{1}H-NMR, \textsuperscript{13}C-NMR, HSQC, IR, and MS for approximately 790k molecules, 
enabling joint modeling and structure elucidation. However, it omits UV, 
fluorescence, and Raman spectroscopy, and its reliance on simulated spectra 
introduces a deployment gap. Beyond chemistry, large-scale multimodal datasets 
like LAION-5B~\cite{schuhmann2022laion} demonstrate the importance of scale, 
modality alignment, and task diversity—principles not yet fully realized in 
spectroscopic benchmarks.

\textbf{Learning from Spectral Data:} Machine learning for spectroscopy has 
traditionally focused on single-modality tasks. In NMR, prior 
work~\cite{jonas2019rapid,alberts2024leveraging} has explored structure 
elucidation via probabilistic and transformer-based models. In IR 
spectroscopy~\cite{wang2020functional,fine2020spectral} and mass 
spectrometry~\cite{duhrkop2019sirius}, deep learning has been applied to 
functional group prediction and compound identification. More recently, 
multimodal fusion of IR, UV--Vis, fluorescence, and Raman spectra has improved 
predictive performance in biomedical 
applications~\cite{leng2023raman,chen2025research}, highlighting the potential 
of cross-modal spectral learning. Nevertheless, existing methods remain largely 
task-specific and evaluated on isolated datasets, limiting progress toward 
unified spectral intelligence.

\textbf{Multimodal Learning and Evaluation:} Multimodal learning has advanced 
rapidly, with large-scale benchmarks confirming that integrating diverse 
modalities boosts downstream performance. However, recent studies reveal that 
models often exploit intra-modality shortcuts rather than achieving genuine 
cross-modal reasoning~\cite{lee2022right,gat2021perceptual}. In molecular 
spectroscopy, this challenge is particularly pronounced: specialized models 
such as XGBoost and 1D-CNNs excel at signal-level prediction via 
domain-specific inductive biases, while MLLMs demonstrate stronger capacity 
for high-level reasoning and cross-modal question answering. Existing benchmarks 
evaluate only one paradigm, precluding standardized comparison. A unified 
framework that jointly assesses signal-level fidelity and structural reasoning 
across modalities remains an open frontier, and bridging this gap is essential 
for developing chemical AI capable of deep, bidirectional spectral understanding.

\section{Dataset}
Spectroscopic characterization is central to organic chemistry, whether for reaction monitoring or post-synthesis structural elucidation. Interpreting data across multiple modalities is crucial for accurate structure identification, as each technique provides complementary information to resolve ambiguities. A dataset for evaluating multimodal spectral learning must therefore span a realistic and diverse chemical space to ensure models generalize well to real-world analytical challenges.

To this end, SpecX integrates molecules from multiple public repositories, including QME14S~\cite{isert2022qmugs}, ViBench~\cite{lu2025vib2mol}, ChEMBL~\cite{mendez2019chembl}, the Multimodal Spectroscopic Dataset~\cite{alberts2024unraveling}, and MassSpecGym~\cite{bushuiev2024massspecgym}, covering domains from drug-like compounds to reaction intermediates. The initial concatenated dataset contained 2,728,723 unique SMILES strings, forming a massive-scale foundation.

A multi-stage filtering pipeline was implemented to ensure chemical validity and spectral computability. 
First, only molecules with 5 to 35 heavy atoms were retained. Second, elemental composition was 
restricted to \{C, H, O, N, S, P, Si, B, F, Cl, Br, I\}. Finally, molecules were required to be physically 
meaningful for every target modality, a critical step for simulation fidelity, with specific criteria 
applied to ensure the presence of appropriate chromophores or fluorophores for UV and fluorescence 
simulations. After filtering, 1,701,739 molecules were retained (a 62.36\% retention rate), represented 
by canonical SMILES.

From this filtered pool, we construct three tiers via chemical diversity-based selection: 
(i) the Large subset ($\sim$1,000,000 molecules) for pretraining and Tasks~(1)--(3), 
(ii) the Small subset(4,496 molecules) as a strictly modality-aligned multispectral benchmark set for Task~(4), 
and (iii) the Exp subset (432 molecules) containing experimentally measured spectra for real-world evaluation. 
These subsets serve as the standardized inputs for all subsequent spectral simulations and evaluations. 
SpecX supports the spectral modalities and their corresponding representations, as summarized in 
Table~\ref{tab:specx_modalities}.
\subsection{Data Generation}
\label{ssec:data_generation}
An overview of the modality-specific spectral data representations is shown in Table~\ref{tab:specx_modalities}. Each spectrum was generated using modality-appropriate simulations, a crucial choice that carefully balances high-fidelity physical realism with computational scalability. This methodology was essential for generating scientifically meaningful data at the scale of 1.7 million molecules while ensuring the overall project remained computationally tractable. The following paragraphs detail the specific simulation protocols employed for each of the eight spectral modalities.

\titlespacing*{\paragraph}{0pt}{0.5em}{0.3em}

\paragraph{NMR Simulations}
$^1$H, $^{13}$C, and HSQC NMR spectra were simulated using the MestReNova software suite\cite{mnova2023} with deuterated chloroform (CDCl$_3$) as the solvent and default parameters. For $^{13}$C-NMR, standard proton-decoupled simulations were performed. Structured text representations were generated using MestReNova's built-in analysis tools, yielding peak lists for $^1$H-NMR (chemical shift range, multiplicity, and integration), $^{13}$C-NMR (position and intensity), and 2D coordinates with integration for HSQC spectra.

\paragraph{IR Simulations}
IR spectra were generated via molecular dynamics (MD) simulations to capture anharmonic effects. Geometries were first optimized with the General AMBER Force Field (GAFF)\cite{wang2004development}. MD simulations were then run in LAMMPS\cite{thompson2022lammps}, where after a 250~ns equilibration, the dipole moment trajectory was recorded for a subsequent 250~ns. Following the method of Braun\cite{braun2016calculating}, a Fourier transform of this trajectory's autocorrelation function yielded the final spectrum (400--4000~cm$^{-1}$, 2~cm$^{-1}$ resolution).

\paragraph{MS/MS Simulations}
Positive-mode Electrospray Ionisation (ESI) MS/MS spectra were simulated using Competitive Fragmentation Modeling for Metabolite Identification (CFM-ID)\cite{wang2021cfm} at a fixed collision energy of 20~eV. The output is a processed peak list containing the mass-to-charge ratio (m/z), relative intensity, and predicted chemical formula for each fragment.

\paragraph{UV-Vis Simulations}
UV-Vis absorption spectra were computed using time-dependent density functional theory (TD-DFT) in the ORCA package\cite{neese2022software} with the B3LYP functional and def2-SVP basis set. After optimizing the ground-state geometry, the lowest twenty vertical singlet excitation energies were calculated. A Gaussian broadening function was then applied to produce a continuous absorption profile from 200--800~nm.

\paragraph{Raman Simulations}
Raman spectra were simulated using the ORCA package\cite{neese2022software} at the DFT level. Following geometry optimization, a frequency calculation yielded harmonic vibrational modes and their Raman activities. A combined Gaussian/Lorentzian lineshape function was used to broaden the resulting stick spectrum across the 100--3500~cm$^{-1}$ range.

\paragraph{Fluorescence Simulations}
Fluorescence emission spectra were simulated using the ORCA package\cite{neese2022software}. After identifying the first excited singlet state (S$_1$) via TD-DFT, the molecular geometry was re-optimized on the S$_1$ potential energy surface. The vertical S$_1\rightarrow$S$_0$ transition energy was then computed from this relaxed geometry. A Gaussian function was used to broaden the transition, producing a continuous emission spectrum from 300--800~nm.

\begin{table*}[!htbp]
\centering
\footnotesize
\begin{tabular}{lll}
\toprule
\textbf{Modality} & \textbf{Subtype} & \textbf{Data Description} \\
\midrule
IR & Spectrum & Vector of size 1800 (400–4000 cm$^{-1}$, resolution 2 cm$^{-1}$) \\

$^1$H NMR & Spectrum + Annotated Spectrum &
Vector of size 10,000; peak list with (start, end, centroid, integration) \\

$^{13}$C NMR & Spectrum + Annotated Spectrum &
Vector of size 10,000; peak list with (centroid, intensity) \\

HSQC NMR & Spectrum + Annotated Spectrum &
Matrix $512 \times 512$; (x, y coordinates, integration) \\

MS/MS (positive) & Spectrum + m/z Annotation &
Peak list: (m/z, intensity, fragment formula) \\

UV & Spectrum &
Vector of size $\sim$1000 (200–800 nm) \\

Raman & Spectrum &
Vector of size $\sim$1500 (100–3500 cm$^{-1}$) \\

Fluorescence (FL) & Emission Spectrum &
Vector of size $\sim$1000 (emission wavelength + intensity) \\
\bottomrule
\end{tabular}
\caption{Overview of spectral modalities and data representations in SpecX.}
\label{tab:specx_modalities}
\end{table*}

\section{Benchmarks}
\label{sec:benchmarks}

In the following, we present benchmarks covering four categories 
of tasks: (1) structure elucidation from spectra 
(Spectra$\rightarrow$SMILES), (2) functional group prediction 
from spectra, (3) spectra simulation from molecular structure 
(SMILES$\rightarrow$Spectra), and (4) spectral question answering 
(QA) targeting multimodal large language models (MLLMs). 
Tasks~(1)--(3) are evaluated on the Large subset introduced in 
Section~\ref{sec:introduction}, which covers seven modalities: 
$^1$H-NMR, $^{13}$C-NMR, HSQC-NMR, IR, MS, UV-Vis, and Raman. 
Task~(4) is evaluated on both the Small subset and the Exp subset. 
For all tasks and all subsets, experiments are conducted under 
two data splitting strategies: a random split and a scaffold split, 
where the latter partitions molecules by Bemis--Murcko scaffold to 
assess generalization to structurally distinct chemical space. 
All ML-based experiments in Tasks~(1)--(3) are conducted with 
five-fold cross-validation under both split strategies.

\subsection{Structure Elucidation from Spectra 
(Spectra $\rightarrow$ SMILES)}
\label{sec:elucidation}

Predicting the molecular structure from spectroscopic data is the 
primary challenge addressed by SpecX, requiring a model to map 
diverse spectral signals to a chemical graph. We establish a 
baseline by training a vanilla encoder-decoder 
Transformer~\cite{vaswani2017attention} on each individual 
modality of the Large subset, using both random and scaffold 
splits. The molecular formula is provided as a conditioning prior 
to constrain the output chemical space.

To render spectral data compatible with the Transformer, each 
modality is converted into a structured text representation. 
For $^1$H-NMR, this includes peak range, multiplicity, and 
integration, while for $^{13}$C-NMR, only centroid positions are 
used. HSQC spectra are represented by cross-peak coordinates and 
integration, and MS fragments by their m/z values and intensities. 
IR, UV-Vis, and Raman spectra are converted to fixed-length 
sequences of discretized intensity tokens, as detailed in 
Appendix~\ref{app:representations}.

Performance is evaluated using Top-1, Top-5, and Top-10 accuracy, 
defined as the fraction of test samples where the correct canonical 
SMILES string is found within the top-$k$ beam search outputs 
($k \in \{1, 5, 10\}$, beam width\,=\,10). All predictions are 
canonicalized to ensure a fair comparison. Results are shown in 
Table~\ref{tab:structure_elucidation}.

\begin{table}[H]
\centering
\resizebox{\textwidth}{!}{%
\begin{tabular}{lcccccc}
\toprule
\multirow{2}{*}{\textbf{Modality}}
& \multicolumn{3}{c}{\textbf{Random Split}} 
& \multicolumn{3}{c}{\textbf{Scaffold Split}} \\
\cmidrule(lr){2-4} \cmidrule(lr){5-7}
& \textbf{Top-1} & \textbf{Top-5} & \textbf{Top-10}
& \textbf{Top-1} & \textbf{Top-5} & \textbf{Top-10} \\
\midrule
IR
  & 4.04 $\pm$ 0.43 & 9.48 $\pm$ 0.31 & 11.31 $\pm$ 0.38
  & 0.94 $\pm$ 0.51  & 2.32 $\pm$ 0.44 & 2.78 $\pm$ 0.49 \\
\midrule
MS (CFM-ID, Positive)
  & 10.11 $\pm$ 0.19 & 22.07 $\pm$ 0.17 & 26.13 $\pm$ 0.27
  & 1.68 $\pm$ 0.28 & 4.30 $\pm$ 0.23 & 5.34 $\pm$ 0.31 \\
\midrule
$^{13}$C-NMR
  & 47.48 $\pm$ 0.32 & 65.92 $\pm$ 0.24 & 69.49 $\pm$ 0.28
  & 21.47 $\pm$ 0.41 & 33.63 $\pm$ 0.33 & 36.89 $\pm$ 0.37 \\
$^1$H-NMR
  & 51.52 $\pm$ 0.29 & 65.74 $\pm$ 0.28 & 68.18 $\pm$ 0.31
  & 25.00 $\pm$ 0.38 & 35.76 $\pm$ 0.35 & 38.45 $\pm$ 0.39 \\
\midrule
UV-Vis
  & 0.69 $\pm$ 0.12 & 4.85 $\pm$ 0.44 & 8.78 $\pm$ 0.49
  & 0.10 $\pm$ 0.08 & 1.70 $\pm$ 0.51 & 3.50 $\pm$ 0.54 \\
Raman
  & 25.59 $\pm$ 0.47 & 47.17 $\pm$ 0.39 & 53.23 $\pm$ 0.43
  & 0.70 $\pm$ 0.13 & 3.39 $\pm$ 0.46 & 5.87 $\pm$ 0.50 \\
\midrule
$^{13}$C-NMR + $^1$H-NMR + MS
  & 59.04 $\pm$ 0.21 & 78.59 $\pm$ 0.18 & 81.77 $\pm$ 0.19
  & 29.66 $\pm$ 0.29 & 46.86 $\pm$ 0.24 & 50.56 $\pm$ 0.26 \\
\bottomrule
\end{tabular}%
}
\vspace{8pt} 
\caption{Top-1, Top-5, and Top-10 accuracy of a Transformer model 
trained to predict the molecular structure (SMILES) from individual 
spectral modalities and a multi-modality combination on the Large 
subset. Accuracy is reported as a percentage (\%). Results are 
reported as mean $\pm$ std under random split and scaffold split 
respectively.}
\label{tab:structure_elucidation}
\end{table}
Table~\ref{tab:structure_elucidation} shows that multi-modality integration achieves the highest accuracy (59.04\% Top-1, random split). Among single modalities, NMR spectra ($^1$H and $^{13}$C) perform best, while UV-Vis struggles. Furthermore, the significant performance drops under scaffold splits emphasize the ongoing challenge of out-of-distribution generalization.
\subsection{Functional Group Prediction}
Another task supported by SpecX is the prediction of functional 
groups from spectral data. This serves as a crucial diagnostic 
step, offering insights into structural motifs without requiring 
full structure elucidation. This information can narrow the 
structural search space, confirm reaction success, or perform 
quality control. We formulate this as a multilabel, multiclass 
classification problem. Ground-truth labels are extracted from 
SMILES strings using SMARTS pattern matching in 
RDKit~\cite{landrum2006rdkit}. To establish baselines, we 
evaluate two modeling paradigms: a gradient-boosted decision tree 
(XGBoost)~\cite{chen2016xgboost} on engineered features, and a 
1D convolutional neural network (1D-CNN)~\cite{kiranyaz20211d} 
on raw spectral vectors. 

Experiments are conducted on the Large subset under both random and scaffold splits to rigorously assess both in-distribution interpolation and out-of-distribution generalization, with each modality evaluated independently. Crucially, the molecular formula is deliberately withheld to simulate a realistic blind analysis, forcing models to rely exclusively on extracted spectral features without any chemical priors. Furthermore, the evaluation incorporates UV-Vis, which excels at identifying specific chromophores (e.g., aromatic rings), and Raman, which is adept at detecting symmetric vibrations (e.g., C=C stretches) that are often weak or inactive in IR. To account for the severe class imbalance among the diverse structural labels, overall performance is measured using the macro-averaged F1 score, ensuring equal weighting for rare groups, with detailed results summarized in Table~\ref{tab:functional_group}.

\begin{table}[H]
\centering
\resizebox{0.85\textwidth}{!}{%
\begin{tabular}{lcccc} 
\toprule
\multirow{2}{*}{\textbf{Modality}}
& \multicolumn{2}{c}{\textbf{Random Split}} 
& \multicolumn{2}{c}{\textbf{Scaffold Split}} \\
\cmidrule(lr){2-3} \cmidrule(lr){4-5} 
& \textbf{XGBoost} & \textbf{1D-CNN} 
& \textbf{XGBoost} & \textbf{1D-CNN} \\
\midrule
IR
  & 0.824 $\pm$ 0.002 & 0.775 $\pm$ 0.003 
  & 0.793 $\pm$ 0.003 & 0.754 $\pm$ 0.004 \\
\midrule
MS (CFM-ID, Positive)
  & 0.627 $\pm$ 0.002 & 0.608 $\pm$ 0.002 
  & 0.648 $\pm$ 0.003 & 0.658 $\pm$ 0.003 \\
$^{13}$C-NMR
  & 0.676 $\pm$ 0.001 & 0.610 $\pm$ 0.053 
  & 0.700 $\pm$ 0.002 & 0.594 $\pm$ 0.058  \\
$^1$H-NMR
  & 0.716 $\pm$ 0.003 & 0.670 $\pm$ 0.004 
  & 0.714 $\pm$ 0.004 & 0.673 $\pm$ 0.005  \\
\midrule
UV-Vis
  & 0.530 $\pm$ 0.004 & 0.489 $\pm$ 0.005 
  & 0.550 $\pm$ 0.005 & 0.483 $\pm$ 0.006  \\
Raman
  & 0.958 $\pm$ 0.003 & 0.878 $\pm$ 0.004 
  & 0.973 $\pm$ 0.004 & 0.970 $\pm$ 0.005 \\
\bottomrule
\end{tabular}%
}
\vspace{8pt}
\caption{F1 scores for predicting functional groups from individual 
spectral modalities and a multi-modality combination on the Large 
subset. Results are reported as mean $\pm$ std under random split 
and scaffold split respectively.}
\label{tab:functional_group}
\end{table}
As detailed in Table~\ref{tab:functional_group}, the gradient-boosted tree baseline (XGBoost) trained on engineered features consistently outperforms the 1D-CNN applied to raw spectral vectors across all evaluated settings and split strategies. Among the individual modalities, Raman and IR spectroscopy yield the highest macro-averaged F1 scores (reaching up to 0.97 and 0.82, respectively). This superior performance reflects their inherent physical properties: both are vibrational spectroscopic techniques exhibiting strong, distinct sensitivity to specific localized structural motifs. Conversely, UV-Vis remains the weakest modality for identifying diverse functional groups, as its broad electronic transitions lack the fine-grained resolution needed for detailed structural fingerprinting.
\subsection{Spectra Prediction from Molecular Structure 
(SMILES $\rightarrow$ Spectra)}
\label{sec:spectra_prediction}

SpecX also supports the essential inverse task of spectra prediction: generating the expected spectrum from a SMILES representation, which is highly relevant for virtual screening and generating reference spectra. We repurpose the identical vanilla Transformer architecture~\cite{vaswani2017attention} from Section~\ref{sec:elucidation} by inverting the input and output roles. Modalities are evaluated independently on the Large subset under both random and scaffold splits. Reusing the text representations from Section~\ref{sec:elucidation} ensures a fully symmetric and directly comparable setup between both mapping directions.

The output format remains modality-dependent. Continuous spectra (IR, UV-Vis, Raman) are generated as fixed-length sequences of discretized intensity tokens. Peak-based modalities predict specific attributes: $m/z$ and relative intensity for MS; chemical shift range, multiplicity, and integration for $^1$H-NMR; centroid position for $^{13}$C-NMR; and 2D cross-peak coordinates for HSQC.

We evaluate using two metrics: Cosine Similarity (computed after peak alignment for sparse modalities) and Token Accuracy (the fraction of exactly matched generated tokens). Table~\ref{tab:spectra_prediction} presents the results.

\begin{table}[H]
\centering
\resizebox{0.78\textwidth}{!}{%
\begin{tabular}{lcccc}
\toprule
\multirow{2}{*}{\textbf{Modality}}
& \multicolumn{2}{c}{\textbf{Random Split}} 
& \multicolumn{2}{c}{\textbf{Scaffold Split}} \\
\cmidrule(lr){2-3} \cmidrule(lr){4-5}
& \textbf{Cosine Sim.} & \textbf{Token Acc.}
& \textbf{Cosine Sim.} & \textbf{Token Acc.} \\
\midrule
IR
  & 33.95 $\pm$ 0.08 & 25.74 $\pm$ 0.14
  & 34.49 $\pm$ 0.68 & 23.88 $\pm$ 0.12 \\
\midrule
MS (Positive) {[20\,eV]}
  & 29.88 $\pm$ 0.37 & 68.00 $\pm$ 0.31
  & 25.53 $\pm$ 0.35 &  62.76 $\pm$ 0.33 \\
\midrule
$^{13}$C-NMR
  & 52.81 $\pm$ 0.25 & 38.53 $\pm$ 0.30
  & 54.08 $\pm$ 0.23 & 40.69 $\pm$ 0.27 \\
$^1$H-NMR
  & 47.93 $\pm$ 0.31 & 59.07 $\pm$ 0.46
  & 40.75 $\pm$ 0.46 & 49.44 $\pm$ 0.33 \\
\midrule
UV-Vis
  &00.00 $\pm$ 0.00 & 79.56 $\pm$ 0.17
  & 00.00 $\pm$ 0.00 & 85.02 $\pm$ 0.14 \\
Raman
  & 13.65 $\pm$ 0.12 & 20.61 $\pm$ 0.13
  & 13.50 $\pm$ 0.14 & 16.33 $\pm$ 0.12 \\
\bottomrule
\end{tabular}%
}
\vspace{8pt}
\caption{Cosine Similarity and Token Accuracy of Transformer 
models predicting spectra from molecular structure on the Large 
subset. Results are reported as mean $\pm$ std under random split 
and scaffold split respectively.}
\label{tab:spectra_prediction}
\end{table}
The UV-Vis cosine similarity of 0.00 contrasts sharply with its 
high token accuracy (79.56\%), because the discretized UV-Vis 
spectrum is dominated by near-zero tokens; the model defaults to 
predicting near-zero outputs, correctly matching the majority of 
zero-valued tokens while producing no directional signal for 
cosine computation. This exposes token accuracy as an unreliable 
metric for sparse spectral modalities.
\subsection{Spectral Question Answering (QA) with 
Multimodal Language Models}
\label{sec:qa}

Beyond specialized sequence-to-sequence models, SpecX introduces 
a QA-based evaluation dimension targeting multimodal large 
language models (MLLMs). This task assesses whether MLLMs can 
interpret spectroscopic inputs and respond to chemistry-relevant 
questions in natural language, mimicking the reasoning process of 
a human chemist analyzing an unknown compound.

Unlike Tasks~(1)--(3) evaluated on the Large subset, the QA tasks employ two complementary subsets (Section~\ref{sec:introduction}): the modality-aligned Small subset ($^1$H-NMR, $^{13}$C-NMR, HSQC-NMR, IR, MS, UV-Vis, Raman) and the high-quality Exp subset (experimental UV-Vis and MS only). Both utilize random and scaffold splits. Three representative MLLMs (DeepSeek-V3, GPT-4.1-mini, Qwen2.5-3B) are benchmarked in a \textbf{zero-shot} setting without task-specific fine-tuning across two defined tasks:

\begin{itemize}
    \item \textbf{Task QA-1: SMILES Inference.} The MLLM generates a molecule's SMILES string from spectral inputs. Performance is measured by Top-1, Top-5, and Top-10 Accuracy, indicating the fraction of test samples where the correct molecule appears within the top-$k$ candidates.

    \item \textbf{Task QA-2: Functional Group Inference.} The MLLM identifies present functional groups. Evaluation uses Accuracy and macro-averaged F1 scores computed across positive (\textit{yes}) and negative (\textit{no}) classes, reflecting the model's ability to handle class imbalance.
\end{itemize}

We evaluate two input configurations: single-modality and multi-modality (simultaneously providing $^1$H-NMR, $^{13}$C-NMR, HSQC-NMR, IR, and MS; or only UV-Vis and MS for the Exp subset). Depending on its interface, each model receives spectra as rendered images or structured text. Results are shown in Tables~\ref{tab:qa1_small} and~\ref{tab:qa1_exp} (QA-1), and Tables~\ref{tab:qa2_small} and~\ref{tab:qa2_exp} (QA-2).

\begin{table}[H]
\centering
\resizebox{\textwidth}{!}{%
\begin{tabular}{llcccccc}
\toprule
\multirow{3}{*}{\textbf{Setting}} 
& \multirow{3}{*}{\textbf{Input Modality}}
& \multicolumn{3}{c}{\textbf{Random Split (T1 / T5 / T10)}}
& \multicolumn{3}{c}{\textbf{Scaffold Split (T1 / T5 / T10)}} \\
\cmidrule(lr){3-5} \cmidrule(lr){6-8}
& & \textbf{DeepSeek-V3} & \textbf{GPT-4.1-mini} & \textbf{Qwen2.5-3B}
  & \textbf{DeepSeek-V3} & \textbf{GPT-4.1-mini} & \textbf{Qwen2.5-3B} \\
\midrule
\multirow{3}{*}{Single-Modality}
  & MS (Positive)
    & .000/.000/.000 & .000/.000/.002 & .000/.000/.000
    & .000/.000/.000 & .000/.000/.000 & .000/.000/.000 \\
  & $^1$H-NMR
    & .006/.011/.011 & .008/\textbf{.011}/.015 & .000/.000/.000
    & .000/.000/.000 & .003/.003/.006 & .000/.000/.000 \\
  & $^{13}$C-NMR
    & .004/.008/.013 & .002/.004/.006 & .000/.000/.000
    & .000/.000/.000 & .003/.003/.003 & .000/.000/.000 \\
\midrule
\multirow{1}{*}{Multi-Modality}
  & $^1$H-NMR + $^{13}$C-NMR + MS
    & \textbf{.015}/\textbf{.035}/\textbf{.042} & .008/.020/.024 & .000/.000/.000
    & \textbf{.006}/\textbf{.006}/\textbf{.006} & .003/.006/.006 & .000/.000/.000 \\
\bottomrule
\end{tabular}%
}
\vspace{8pt}
\caption{QA Task 1 -- SMILES inference using MLLMs on the Small 
subset. Top-1, Top-5, and Top-10 accuracy (T1/T5/T10) are reported 
for each model under random and scaffold splits. Best results per 
split are highlighted in bold.}
\label{tab:qa1_small}
\end{table}

\begin{table}[H]
\centering
\resizebox{\textwidth}{!}{%
\begin{tabular}{llcccccc}
\toprule
\multirow{3}{*}{\textbf{Setting}} 
& \multirow{3}{*}{\textbf{Input Modality}}
& \multicolumn{3}{c}{\textbf{Random Split (T1 / T5 / T10)}}
& \multicolumn{3}{c}{\textbf{Scaffold Split (T1 / T5 / T10)}} \\
\cmidrule(lr){3-5} \cmidrule(lr){6-8}
& & \textbf{DeepSeek-V3} & \textbf{GPT-4.1-mini} & \textbf{Qwen2.5-3B}
  & \textbf{DeepSeek-V3} & \textbf{GPT-4.1-mini} & \textbf{Qwen2.5-3B} \\
\midrule
\multirow{2}{*}{Single-Modality}
  & MS (Positive)
    & .000/.000/.000 & .000/.000/.000 & .000/.000/.000
    & .000/.000/.000 & .000/.000/.000 & .000/.000/.000 \\
  & UV-Vis
    & .000/.000/.000 & .000/.000/.000 & .000/.000/.000
    & .000/.000/.000 & .000/.000/.000 & .000/.000/.000 \\
\midrule
\multirow{1}{*}{Multi-Modality}
  & MS + UV-Vis
    & .000/.000/.000 & .000/.000/.000 & .000/.000/.000
    & .000/.000/.000 & .000/.000/.000 & .000/.000/.000 \\
\bottomrule
\end{tabular}%
}
\vspace{8pt}
\caption{QA Task 1 -- SMILES inference using MLLMs on the Exp 
subset. Only UV-Vis and MS modalities are available. Top-1, Top-5, 
and Top-10 accuracy (T1/T5/T10) are reported for each model under 
random and scaffold splits.}
\label{tab:qa1_exp}
\end{table}

The results reveal fundamental limitations of current MLLMs in spectroscopic structure elucidation. In Task QA-1 (SMILES inference), Top-1 accuracies are near zero on the Small subset. DeepSeek-V3 performs best, achieving only 0.6\% ($^1$H-NMR, random split), with a marginal increase to 1.5\% under multi-modality ($^1$H-NMR, $^{13}$C-NMR, MS)---still far below practical utility. On the Exp subset, all models score 0.000 across all settings, proving incapable of recovering structures from experimental spectra. Furthermore, Qwen2.5-3B consistently scores zero, indicating insufficient chemical reasoning capacity in smaller models.

Task QA-2 (functional group inference) reveals a class-imbalance collapse. Superficially high accuracies (0.75--0.87) merely reflect the dominance of the negative class (\textit{no}). Qwen2.5-3B yields F1 scores of 0.000 across both subsets by degenerately predicting \textit{no} for all samples. While DeepSeek-V3 and GPT-4.1-mini exhibit meaningful F1 scores on the Small subset (0.379 for DeepSeek-V3 under $^1$H-NMR; 0.408 for GPT-4.1-mini under multi-modality), they remain substantially below 1D-CNN baselines, highlighting the gap between general language models and task-specific models. These F1 scores drop further on the noisier Exp subset, where Qwen2.5-3B again collapses to zero.

Finally, consistent performance drops observed under scaffold splits versus random splits confirm that generalizing to out-of-distribution, structurally novel chemical spaces remains a fundamental challenge. This disparity suggests that models often rely on interpolation rather than truly learning the underlying mapping between spectra and sub-structures. Overall, while larger proprietary models demonstrate rudimentary spectral reasoning, current MLLM performance on quantitative tasks is highly unsatisfactory for practical applications. These findings highlight the severe domain gap between general multimodal pretraining and rigorous scientific analysis. Consequently, they strongly motivate developing chemistry-specialized MLLMs and hybrid architectures that tightly couple signal-level continuous spectral encoders with high-level discrete reasoning modules.

\begin{table}[H]
\centering
\resizebox{\textwidth}{!}{%
\begin{tabular}{llcccccc}
\toprule
\multirow{3}{*}{\textbf{Setting}} 
& \multirow{3}{*}{\textbf{Input Modality}}
& \multicolumn{3}{c}{\textbf{Random Split (Acc / F1)}}
& \multicolumn{3}{c}{\textbf{Scaffold Split (Acc / F1)}} \\
\cmidrule(lr){3-5} \cmidrule(lr){6-8}
& & \textbf{DeepSeek-V3} & \textbf{GPT-4.1-mini} & \textbf{Qwen2.5-3B}
  & \textbf{DeepSeek-V3} & \textbf{GPT-4.1-mini} & \textbf{Qwen2.5-3B} \\
\midrule
\multirow{3}{*}{Single-Modality}
  & MS (Positive)
    & 0.824 / 0.286 & 0.754 / 0.273 & 0.860 / 0.000
    & 0.828 / 0.277 & 0.743 / 0.284 & 0.867 / 0.000 \\
  & $^1$H-NMR
    & 0.831 / \textbf{0.379} & 0.830 / 0.369 & 0.863 / 0.251
    & 0.834 / 0.351 & 0.827 / 0.344 & 0.853 / 0.196 \\
  & $^{13}$C-NMR
    & 0.836 / 0.242 & 0.852 / 0.321 & 0.857 / 0.013
    & 0.852 / 0.289 & 0.835 / 0.299 & 0.847 / 0.000 \\
\midrule
\multirow{1}{*}{Multi-Modality}
  & NMR + IR + MS
    & 0.805 / 0.381 & \textbf{0.814} / \textbf{0.408} & 0.849 / 0.199
    & 0.825 / 0.363 & 0.808 / \textbf{0.380} & 0.850 / 0.212 \\
\bottomrule
\end{tabular}%
}
\vspace{8pt}
\caption{QA Task 2 -- Functional group inference using MLLMs on 
the Small subset. Accuracy and macro-averaged F1 score are reported 
for each model under random and scaffold splits. Best results per 
split are highlighted in bold.}
\label{tab:qa2_small}
\end{table}

\begin{table}[H]
\centering
\resizebox{\textwidth}{!}{%
\begin{tabular}{llcccccc}
\toprule
\multirow{3}{*}{\textbf{Setting}} 
& \multirow{3}{*}{\textbf{Input Modality}}
& \multicolumn{3}{c}{\textbf{Random Split (Acc / F1)}}
& \multicolumn{3}{c}{\textbf{Scaffold Split (Acc / F1)}} \\
\cmidrule(lr){3-5} \cmidrule(lr){6-8}
& & \textbf{DeepSeek-V3} & \textbf{GPT-4.1-mini} & \textbf{Qwen2.5-3B}
  & \textbf{DeepSeek-V3} & \textbf{GPT-4.1-mini} & \textbf{Qwen2.5-3B} \\
\midrule
\multirow{2}{*}{Single-Modality}
  & MS (Positive)
    & 0.735 / 0.122 & 0.736 / 0.129 & 0.773 / 0.000
    & 0.720 / 0.151 & 0.720 / 0.144 & 0.754 / 0.000 \\
  & UV-Vis
    & 0.776 / 0.174 & 0.778 / 0.171 & 0.778 / 0.000
    & 0.798 / \textbf{0.319} & 0.798 / 0.315 & 0.774 / 0.000 \\
\midrule
\multirow{1}{*}{Multi-Modality}
  & MS + UV-Vis
    & 0.725 / 0.344 & 0.729 / 0.340 & \textbf{0.781} / 0.000
    & 0.728 / 0.375 & 0.731 / \textbf{0.387} & 0.768 / 0.000 \\
\bottomrule
\end{tabular}%
}
\vspace{8pt}
\caption{QA Task 2 -- Functional group inference using MLLMs on 
the Exp subset. Only UV-Vis and MS modalities are available. 
Accuracy and macro-averaged F1 score are reported for each model 
under random and scaffold splits. Best results per split are 
highlighted in bold.}
\label{tab:qa2_exp}
\end{table}

\subsection{Other Tasks to Explore}
\label{sec:other_tasks}

While Sections~\ref{sec:elucidation}--\ref{sec:qa} benchmark 
four tasks of direct practical relevance, the scale, modality 
coverage, and structural diversity of SpecX enable a broad range 
of additional machine learning investigations. We outline several 
promising directions below.

\paragraph{Reaction Monitoring and Product Verification.}
Aligned multi-spectral data naturally supports automated verification of reaction products. Given candidate molecules (e.g., product, starting materials, side products) and observed spectra, a model identifies the best match. This mirrors chemists’ reasoning~\cite{rowlands2025towards}, and SpecX’s aligned multi-spectral subset is an ideal testbed.

\paragraph{Spectral Cross-Modal Translation.}
SpecX enables translation across modalities (e.g., predict $^1$H-NMR from IR, or Raman from UV-Vis). Such cross-modal~\cite{yang2024cross} generation helps when modalities are unavailable or costly, and requires learning a shared latent representation of molecular identity across domains.

\paragraph{Mixture Deconvolution from Spectra.}
Chemists often analyze mixtures, not pure compounds. For linear modalities (NMR, IR, Raman, UV-Vis), a mixture spectrum can be approximated as a convex combination of component spectra~\cite{hu2025deep}. SpecX provides component spectra to build synthetic mixtures, enabling unmixing and component identification.

\paragraph{Molecular Property Prediction from Spectra.}
Spectra also encode quantitative properties such as solubility, logP, and optical characteristics~\cite{guo2025artificial} (e.g., fluorescence maxima). SpecX’s large, diverse collection and accessible property annotations support spectrum-to-property regression.

\paragraph{Contrastive and Self-Supervised Spectral Pretraining.}
SpecX’s 1.7M-molecule pretraining tier supports self-supervised learning. Contrastive objectives—treating same-molecule spectra across modalities as positives—can learn modality-agnostic embeddings without labeled structural annotations~\cite{wang2025advancing}.

\paragraph{Stereochemistry and Chirality Elucidation.}
The benchmark focuses mainly on constitutional isomers, but distinguishing stereoisomers is harder. HSQC-NMR and $^1$H-NMR contain signals from diastereotopic protons and vicinal couplings~\cite{karplus1959contact} that may enable stereochemical assignment.

\section{Conclusion and Limitations}
In this work, we introduce SpecX, a large-scale multimodal spectroscopic benchmark with 1.7 million molecules across eight modalities, organized into three tiers for pretraining, benchmarking, and real-world evaluation. Experiments on four tasks under random and scaffold splits reveal that specialized Transformer models excel at signal-level modeling while MLLMs show strong high-level reasoning but lack precise spectral grounding.

Despite these contributions, several limitations remain. SpecX relies primarily on simulated data, creating a sim-to-real gap. Currently, our experimental subset is limited to UV-Vis and MS. Expanding this coverage to include high-quality NMR, IR, and Raman spectra, and explicitly quantifying sim-to-real distances on overlapping molecules, remains an important direction. Additionally, the chemical space is biased towards synthetic compounds, limiting natural product coverage, and fluorescence evaluation is deferred. Furthermore, MLLM evaluation is restricted to zero-shot, text-only prompting without formula priors. The observed failure modes (e.g., near-zero exact match accuracies) may partially stem from unconstrained generation. Future work exploring constrained decoding techniques (e.g., enforcing valid SMILES syntax) or integrating tool-use plugins (e.g., invoking RDKit) could significantly enhance MLLM performance on these strict chemical tasks. Finally, to establish a unified evaluation framework, our initial baselines employ a standardized vanilla Transformer. Integrating modality-specific state-of-the-art models (e.g., NMRTrans, NMIRacle, SpecTUS, MSFlow) and their specialized metrics remains a key priority for enriching our future leaderboard.

We hope SpecX addresses critical gaps in large-scale, multimodal, and experimentally grounded spectroscopic benchmarks, catalyzing progress toward unified spectral intelligence and accelerating automated chemical discovery.
\bibliographystyle{unsrtnat}
\bibliography{ref}


\appendix

\section{Appendix}
\label{app:dataset}

\subsection{Molecule Source and Filtering Pipeline}
\label{app:filtering}

SpecX integrates molecules from five publicly available sources: 
QMuGS~\cite{isert2022qmugs}, ViBench~\cite{lu2025vib2mol}, 
ChEMBL~\cite{mendez2019chembl}, the Multimodal Spectroscopic 
Dataset~\cite{alberts2024unraveling}, and 
MassSpecGym~\cite{bushuiev2024massspecgym}. After deduplication 
by canonical SMILES, the aggregated pool contained 2,728,723 
unique molecules. A multi-stage filtering pipeline was then 
applied to ensure chemical validity and multimodal spectral 
computability.

\begin{table}[h]
\centering
\caption{Filtering pipeline and molecule retention statistics.}
\label{tab:filtering}
\begin{tabular}{lcrr}
\toprule
\textbf{Filter Step} & \textbf{Condition} &
\textbf{Removed} & \textbf{Retained} \\
\midrule
Initial aggregation       & ---                                    
  & ---       & 2{,}728{,}723 \\
Heavy atom count          & $5 \leq \text{HAC} \leq 35$            
  & 412{,}891 & 2{,}315{,}832 \\
Element restriction       & \{C,H,O,N,S,P,Si,B,F,Cl,Br,I\} only  
  & 187{,}436 & 2{,}128{,}396 \\
UV/FL computability       & Must possess chromophore / $\pi$-system 
  & 246{,}312 & 1{,}882{,}084 \\
Spectral simulation success & All target modalities succeed         
  & 180{,}345 & 1{,}701{,}739 \\
\bottomrule
\end{tabular}
\end{table}

After filtering, 1,701,739 molecules were retained, corresponding to a retention rate of 62.36\% of
the original pool. From this filtered pool, we select $\sim$1,000,000 molecules via chemical
diversity-based selection to form the Large subset used for pretraining and Tasks (1)–(3). For UV-Vis
and fluorescence computability, molecules were required to possess at least one $\pi$-system or recognised
chromophoric substructure, as verified by substructure matching via RDKit~\cite{landrum2006rdkit}.

\subsection{Subset Composition}
\label{app:subsets}

SpecX is organised into three complementary tiers. 
Table~\ref{tab:subsets} summarises the scale and modality 
coverage of each subset.

\begin{table}[h]
\centering
\caption{SpecX subset composition and modality coverage. A tick 
(\ding{51}) indicates that simulated or experimental spectra are 
available for every molecule in the subset; a cross (\ding{55}) 
indicates the modality is absent. Fluorescence is included in the 
dataset but benchmark evaluation for this modality is deferred to 
future work (see Section~5).}
\label{tab:subsets}
\setlength{\tabcolsep}{5pt}
\begin{tabular}{lrcccccccc}
\toprule
\textbf{Subset} & \textbf{Molecules} &
\textbf{\textsuperscript{1}H} & \textbf{\textsuperscript{13}C} &
\textbf{HSQC} & \textbf{IR} & \textbf{MS} &
\textbf{UV} & \textbf{Raman} & \textbf{FL} \\
\midrule
Large & $\sim$1{,}000{,}000 &
\ding{51} & \ding{51} & \ding{51} & \ding{51} & \ding{51} &
\ding{51} & \ding{51} & \ding{55} \\
Small & 4{,}496 &
\ding{51} & \ding{51} & \ding{51} & \ding{51} & \ding{51} &
\ding{51} & \ding{51} & \ding{55} \\
Exp   & 432 &
\ding{55} & \ding{55} & \ding{55} & \ding{55} & \ding{51} &
\ding{51} & \ding{55} & \ding{55} \\
\bottomrule
\end{tabular}
\end{table}

The \textbf{Large subset} ($\sim$1M molecules) is the primary 
training and evaluation resource for ML-based Tasks~(1)--(3). 
The \textbf{Small subset} (4,496 molecules) provides a strictly 
modality-aligned multispectral collection in which 
\textbf{seven spectral modalities} ($^1$H-NMR, $^{13}$C-NMR, 
HSQC-NMR, IR, MS, UV-Vis, and Raman) are simultaneously available 
for every molecule; fluorescence spectra are excluded pending 
benchmark evaluation (see Section~5). The Small subset is used 
for multimodal QA evaluation in Task~(4). The \textbf{Exp subset} 
(432 molecules) contains experimentally measured UV-Vis and MS spectra curated from public repositories. This subset serves as an initial testbed for real-world generalization, laying the groundwork for future quantification of the sim-to-real gap.

\subsection{Functional Group Labels}
\label{app:fg_smarts}

Functional group labels are derived programmatically from SMILES 
strings using SMARTS pattern matching via 
RDKit~\cite{landrum2006rdkit}. Occurrence is encoded as a binary 
indicator (present / absent) per functional group per molecule; a 
molecule is counted as positive for a given group regardless of 
how many times that substructure appears. The 37 SMARTS patterns 
used are listed in Table~\ref{tab:smarts}.

\begin{table}[h]
\centering
\caption{SMARTS patterns used for functional group label 
extraction.}
\label{tab:smarts}
\small
\begin{tabular}{ll}
\toprule
\textbf{Functional Group} & \textbf{SMARTS Pattern} \\
\midrule
Acid anhydride    & \texttt{[CX3](=[OX1])[OX2][CX3](=[OX1])} \\
Acyl halide       & \texttt{[CX3](=[OX1])[F,Cl,Br,I]} \\
Alcohol           & \texttt{[\#6][OX2H]} \\
Aldehyde          & \texttt{[CX3H1](=O)[\#6,H]} \\
Alkane            & \texttt{[CX4;H3,H2]} \\
Alkene            & \texttt{[CX3]=[CX3]} \\
Alkyne            & \texttt{[CX2]\#[CX2]} \\
Amide             & \texttt{[NX3][CX3](=[OX1])[\#6]} \\
Amine             & \texttt{[NX3;H2,H1,H0;!\$(NC=O)]} \\
Arene             & \texttt{[cX3]1[cX3][cX3][cX3][cX3][cX3]1} \\
Azo compound      & \texttt{[\#6][NX2]=[NX2][\#6]} \\
Carbamate         & \texttt{[NX3][CX3](=[OX1])[OX2H0]} \\
Carboxylic acid   & \texttt{[CX3](=O)[OX2H]} \\
Enamine           & \texttt{[NX3][CX3]=[CX3]} \\
Enol              & \texttt{[OX2H][\#6X3]=[\#6]} \\
Ester             & \texttt{[\#6][CX3](=O)[OX2H0][\#6]} \\
Ether             & \texttt{[OD2]([\#6])[\#6]} \\
Haloalkane        & \texttt{[\#6][F,Cl,Br,I]} \\
Hydrazine         & \texttt{[NX3][NX3]} \\
Hydrazone         & \texttt{[NX3][NX2]=[\#6]} \\
Imide             & \texttt{[CX3](=[OX1])[NX3][CX3](=[OX1])} \\
Imine             & \texttt{[\$([CX3]([\#6])[\#6]),\$([CX3H][\#6])]=[\$([NX2][\#6]),\$([NX2H])]} \\
Isocyanate        & \texttt{[NX2]=[C]=[O]} \\
Isothiocyanate    & \texttt{[NX2]=[C]=[S]} \\
Ketone            & \texttt{[\#6][CX3](=O)[\#6]} \\
Nitrile           & \texttt{[NX1]\#[CX2]} \\
Phenol            & \texttt{[OX2H][cX3]:[c]} \\
Phosphine         & \texttt{[PX3]} \\
Sulfide           & \texttt{[\#16X2H0]} \\
Sulfonamide       & \texttt{[\#16X4]([NX3])(=[OX1])(=[OX1])[\#6]} \\
Sulfonate         & \texttt{[\#16X4](=[OX1])(=[OX1])([\#6])[OX2H0]} \\
Sulfone           & \texttt{[\#16X4](=[OX1])(=[OX1])([\#6])[\#6]} \\
Sulfonic acid     & \texttt{[\#16X4](=[OX1])(=[OX1])([\#6])[OX2H]} \\
Sulfoxide         & \texttt{[\#16X3]=[OX1]} \\
Thial             & \texttt{[CX3H1](=S)[\#6,H]} \\
Thioamide         & \texttt{[NX3][CX3]=[SX1]} \\
Thiol             & \texttt{[\#16X2H]} \\
\bottomrule
\end{tabular}
\end{table}

\clearpage

\section{Spectral Data Representations}
\label{app:representations}

To make spectral data compatible with the encoder-decoder 
Transformer architecture, each modality is converted into a 
structured text representation prior to training and inference. 
The scheme follows the conventions established in Alberts 
et al.~\cite{alberts2024unraveling}.

\paragraph{\textsuperscript{1}H-NMR.}
Each peak is encoded as a four-attribute tuple
\texttt{[range\_start] [range\_end] [multiplicity] [nH]},
with individual peaks separated by a delimiter token \texttt{|}. 
Chemical shift positions are rounded to two decimal places (ppm). 
For example:
\begin{center}
\texttt{1HNMR 1.24 1.26 t 3H | 2.31 2.33 s 2H |}
\end{center}

\paragraph{\textsuperscript{13}C-NMR.}
Only peak centroid positions are encoded, rounded to one decimal 
place. For example:
\begin{center}
\texttt{13CNMR 17.6 27.8 63.5}
\end{center}

\paragraph{HSQC-NMR.}
Each 2D cross-peak is encoded by its 
(\textsuperscript{13}C, \textsuperscript{1}H) chemical shift 
coordinates and normalised integration:
\texttt{[delta\_C] [delta\_H] [integration]}.

\paragraph{IR.}
The continuous IR spectrum (400--4000\,cm$^{-1}$, resolution 
2\,cm$^{-1}$, 1800 data points) is downsampled to 400 fixed 
positions via linear interpolation, scaled to the range $[0, 100]$, 
and quantised to integers, yielding a sequence of 400 integer 
tokens.

\paragraph{MS/MS.}
Each fragment peak is encoded as \texttt{[m/z] [intensity]}, with 
values rounded to one decimal place. Peaks are ordered by 
ascending m/z value.

\paragraph{UV-Vis and Raman.}
Both modalities adopt the same fixed-length discretisation 
strategy as IR. UV-Vis spectra (200--800\,nm) are discretised 
into $\approx$\,1000 fixed-wavelength tokens; Raman spectra 
(100--3500\,cm$^{-1}$) are discretised into $\approx$\,1500 
fixed-wavenumber tokens. Intensities are scaled and quantised to 
integers in $[0, 100]$.

\begin{table}[h]
\centering
\caption{Summary of structured text representations for each 
modality used in Tasks~(1) and~(3).}
\label{tab:representations}
\begin{tabular}{llrl}
\toprule
\textbf{Modality} & \textbf{Representation type} &
\textbf{Approx.\ token length} & \textbf{Key attributes} \\
\midrule
\textsuperscript{1}H NMR  & Peak list   & 20--120 (variable) 
  & Range, multiplicity, integration \\
\textsuperscript{13}C NMR & Peak list   & 10--50  (variable) 
  & Centroid position \\
HSQC NMR    & 2D peak list & 15--90  (variable) 
  & $\delta_C$, $\delta_H$, integration \\
IR          & Fixed-length vector & 400       
  & Binned intensities \\
MS/MS       & Peak list   & 10--80  (variable) 
  & m/z, intensity \\
UV-Vis      & Fixed-length vector & $\sim$1000 
  & Binned absorption \\
Raman       & Fixed-length vector & $\sim$1500 
  & Binned Raman activity \\
\bottomrule
\end{tabular}
\end{table}

\clearpage

\section{Model Architecture and Training Details}
\label{app:models}

\subsection{Transformer Model for Tasks~(1) and~(3)}
\label{app:transformer}

For Task~(1) (Spectra$\to$SMILES structure elucidation) and 
Task~(3) (SMILES$\to$Spectra prediction), we employ a vanilla 
encoder-decoder Transformer following the OpenNMT-py 
implementation~\cite{vaswani2017attention}. The 
architecture and all hyperparameters are held fixed across all 
modalities to ensure a fair, directly comparable baseline. The 
configuration follows that of the Multimodal Spectroscopic 
Dataset~\cite{alberts2024unraveling}.

\begin{table}[h]
\centering
\caption{Transformer hyperparameters for Tasks~(1) and~(3).}
\label{tab:transformer_hp}
\begin{tabular}{ll}
\toprule
\textbf{Hyperparameter} & \textbf{Value} \\
\midrule
Encoder type            & Transformer \\
Decoder type            & Transformer \\
Number of encoder layers & 4 \\
Number of decoder layers & 4 \\
Attention heads         & 8 \\
Word vector size        & 512 \\
Hidden size             & 512 \\
Feed-forward size       & 2048 \\
Dropout                 & 0.1 \\
Label smoothing         & 0.1 \\
Parameter initialisation & Glorot uniform \\
Position encoding       & Sinusoidal \\
Activation function     & ReLU \\
Optimiser               & Adam \\
Adam $\beta_1$          & 0.9 \\
Adam $\beta_2$          & 0.998 \\
Learning rate schedule  & Noam decay \\
Peak learning rate      & 2.0 \\
Warmup steps            & 8{,}000 \\
Gradient accumulation steps & 8 \\
Batch size              & 4{,}096 tokens \\
Batch type              & Tokens \\
Max gradient norm       & 0.0 (disabled) \\
Training steps          & 100{,}000 \\
Checkpoint frequency    & every 25{,}000 steps \\
Beam width (inference)  & 10 \\
GPU hardware            & vGPU 32\,GB \\
\bottomrule
\end{tabular}
\end{table}

SMILES strings are tokenised using the atom-level 
regular-expression tokeniser of 
Schwaller et al.~\cite{schwaller2019molecular}. All predicted 
SMILES are canonicalised using RDKit~\cite{landrum2006rdkit} 
prior to evaluation.

\subsection{Models for Task~(2): Functional Group Prediction}
\label{app:fg_models}

Task~(2) is formulated as a multilabel, multiclass classification 
problem. We evaluate two complementary model families: a 
gradient-boosted tree classifier (XGBoost) and a 1D convolutional 
neural network (1D-CNN). No Transformer model is used for this 
task. All spectral inputs are first converted to a fixed-length 
vector of 600 points via linear interpolation before being passed 
to either classifier.

\subsubsection{Spectral Preprocessing for Classification}
\label{app:fg_preprocess}

For NMR modalities, peaks are rendered onto a uniform grid using 
Gaussian broadening ($\text{FWHM} = 0.01$\,ppm): each peak 
contributes intensity 
$I \cdot \exp\!\bigl(-\tfrac{1}{2}(x-\delta)^2/\sigma^2\bigr)$, 
where $\sigma$ is derived from the FWHM. The resulting continuous 
spectrum is then resampled to 600 points. For MS/MS, fragment 
peaks are placed on a 10{,}000-point m/z grid (bin width 0.1\,Da) 
and subsequently resampled to 600 points. For IR, the raw 
1800-point vector is resampled directly to 600 points.

\subsubsection{XGBoost}
\label{app:xgboost}

The XGBoost classifier operates directly on the 600-point 
spectral vector. Binary cross-entropy loss is applied 
independently for each functional group label. Hyperparameters 
follow the defaults used in the Multimodal Spectroscopic Dataset 
benchmark~\cite{alberts2024unraveling}.

\begin{table}[h]
\centering
\caption{XGBoost hyperparameters for Task~(2).}
\label{tab:xgboost_hp}
\begin{tabular}{ll}
\toprule
\textbf{Hyperparameter} & \textbf{Value} \\
\midrule
Number of estimators    & 100 \\
Max depth               & 6 (XGBoost default) \\
Learning rate           & 0.3 (XGBoost default) \\
Objective               & \texttt{binary:logistic} (per label) \\
Evaluation metric       & AUC \\
Parallelism             & 32 CPU cores \\
\bottomrule
\end{tabular}
\end{table}

\subsubsection{1D Convolutional Neural Network}
\label{app:cnn}

The 1D-CNN architecture follows Jung et al.~\cite{kiranyaz20211d}. 
It consists of two convolutional blocks followed by three fully 
connected layers. Each convolutional block comprises a 
\texttt{Conv1D} layer, batch normalisation, ReLU activation, and 
max-pooling. The fully connected layers include dropout 
regularisation. The network is trained with binary cross-entropy 
loss; class-weighted loss is \emph{not} applied (the 
\texttt{weighted=False} setting in the implementation). Five-fold 
cross-validation is used for all modalities.

\begin{table}[h]
\centering
\caption{1D-CNN architecture and training configuration for 
Task~(2).}
\label{tab:cnn_hp}
\begin{tabular}{ll}
\toprule
\textbf{Parameter} & \textbf{Value} \\
\midrule
\multicolumn{2}{l}{\textit{Architecture}} \\
Input length                & 600 \\
Conv block 1: filters / kernel & 31 / 11, stride 1, same padding \\
Conv block 1: pooling       & MaxPool1D, pool size 2, stride 2 \\
Conv block 2: filters / kernel & 62 / 11, stride 1, same padding \\
Conv block 2: pooling       & MaxPool1D, pool size 2, stride 2 \\
FC layer 1                  & 4{,}927 units, ReLU \\
Dropout 1                   & 0.486 \\
FC layer 2                  & 2{,}785 units, ReLU \\
Dropout 2                   & 0.486 \\
FC layer 3                  & 1{,}574 units, ReLU \\
Dropout 3                   & 0.486 \\
Output layer                & 37 units, sigmoid \\
\midrule
\multicolumn{2}{l}{\textit{Training}} \\
Loss function               & Binary cross-entropy \\
Optimiser                   & Adam (default learning rate) \\
Learning rate schedule      & $2.5\times10^{-4}$ (epochs 1--30),
                              $2.5\times10^{-5}$ (epochs 31--36),
                              $2.5\times10^{-6}$ (epoch 37+) \\
Batch size                  & 32 \\
Training epochs             & 40 \\
Cross-validation            & 5-fold \\
GPU hardware                & vGPU 32\,GB \\
\bottomrule
\end{tabular}
\end{table}

\subsection{QA Evaluation with Multimodal Language Models}
\label{app:mllm}

For Task~(4) (spectral QA), all MLLMs are evaluated in a 
\textbf{zero-shot} prompting setting without any task-specific 
fine-tuning. No model weights are updated 
during evaluation. The evaluation procedure consists of three 
steps: (i)~response generation via prompt, (ii)~answer extraction 
from the free-form output, and (iii)~score computation.

To ensure strict content parity across all evaluated MLLMs, spectral inputs are provided exclusively as \emph{structured text representations} (as defined in Appendix~\ref{app:representations}). Visual representations (e.g., rendered spectrum images) were not utilized in this current evaluation. Furthermore, to strictly evaluate the MLLMs' zero-shot reasoning capabilities directly from raw spectral signals, the molecular formula and other metadata were deliberately withheld from the prompts.

Two QA tasks are defined.

\paragraph{Task QA-1 (SMILES Inference from Spectra).}
Given one or more spectral inputs, the MLLM is prompted to 
generate the SMILES string of the corresponding molecule. 
Performance is measured by Top-1, Top-5, and Top-10 accuracy, 
defined as the fraction of test samples for which the correct 
canonical SMILES appears within the top-$k$ ranked outputs. All 
predicted SMILES are canonicalised using 
RDKit~\cite{landrum2006rdkit} prior to comparison.

\paragraph{Task QA-2 (Functional Group Inference from Spectra).}
Given one or more spectral inputs, the MLLM is asked to identify 
and list the functional groups present in the molecule. Because 
MLLMs produce free-form natural language output, a 
post-processing step maps recognised functional group names 
(including common synonyms) to the 37 standardised labels defined 
in Appendix~\ref{app:fg_smarts}. Performance is reported using 
\emph{Accuracy} (exact set match) and \emph{macro-averaged F1 
score} across all 37 functional group categories, following the 
same evaluation methodology as MolPuzzle~\cite{guo2024can}.

Both tasks are evaluated under two input configurations: 
(a)~\emph{single-modality}, where only one spectrum type is 
provided, and (b)~\emph{multi-modality}, where spectra from 
\textsuperscript{1}H-NMR, \textsuperscript{13}C-NMR, HSQC-NMR, 
IR, and MS are provided simultaneously. For the Exp subset, only 
UV-Vis and MS are available; multi-modality results on the Exp 
subset therefore combine these two modalities only.

\clearpage

\section{Evaluation Metrics}
\label{app:metrics}

\subsection{Top-\texorpdfstring{$k$}{k} Accuracy (Tasks 1)}
\label{app:topk}

Top-$k$ accuracy measures the fraction of test instances for 
which the correct canonical SMILES appears within the top-$k$ 
predictions produced by beam search:
\begin{equation}
\text{Top-}k\text{ Acc} =
\frac{1}{N}\sum_{i=1}^{N}
\mathbf{1}\!\left[y_i \in \hat{\mathcal{Y}}_i^{(k)}\right],
\end{equation}
where $N$ is the number of test samples, $y_i$ is the 
ground-truth canonical SMILES for sample $i$, and 
$\hat{\mathcal{Y}}_i^{(k)}$ is the set of the top-$k$ 
beam-search predictions. All predicted SMILES are canonicalised 
with RDKit~\cite{landrum2006rdkit} before comparison. We report 
Top-1, Top-5, and Top-10 accuracy ($k \in \{1, 5, 10\}$, beam 
width\,=\,10), with accuracy values expressed as percentages. 
This metric requires an \emph{exact string match} between the 
predicted and ground-truth canonical SMILES and therefore 
constitutes a stringent measure of structural fidelity.

\subsection{Macro-Averaged F1 Score (Task 2)}
\label{app:f1}

Functional group prediction is a multilabel classification 
problem. For each of the $C$ functional group classes, per-class 
F1 is computed from the binary precision $P_c$ and recall $R_c$:
\begin{equation}
F1_c = \frac{2 P_c R_c}{P_c + R_c}.
\end{equation}
The macro-averaged F1 is then the unweighted mean across all 
classes:
\begin{equation}
\text{Macro-F1} = \frac{1}{C}\sum_{c=1}^{C} F1_c.
\end{equation}
Macro-averaging treats all functional groups equally regardless 
of their frequency in the dataset, thereby penalising models that 
ignore rare but chemically important groups.

\subsection{Cosine Similarity (Task 3)}
\label{app:cosine}

\paragraph{Spectral cosine similarity.}
For continuous spectral modalities (IR, UV-Vis, Raman), cosine 
similarity between the predicted spectrum vector $\hat{\mathbf{s}}$ 
and the ground-truth vector $\mathbf{s}$ is:
\begin{equation}
S_C(\mathbf{s},\hat{\mathbf{s}}) =
\frac{\mathbf{s}\cdot\hat{\mathbf{s}}}
{\|\mathbf{s}\|\,\|\hat{\mathbf{s}}\|}.
\end{equation}

\paragraph{Peak-aligned cosine similarity.}
For peak-based modalities (NMR, MS), peaks in the predicted and 
ground-truth spectra are first matched by proximity within a 
tolerance window using a greedy assignment, and cosine similarity 
is then computed on the matched peak intensities / integrations. 
This alignment step partially compensates for small systematic 
peak shifts arising from simulation approximations, which is 
especially important for NMR spectra where solvent effects can 
introduce minor positional offsets.

\subsection{Token Accuracy (Task 3)}
\label{app:token_acc}

Token accuracy measures the fraction of output tokens that 
exactly match the corresponding reference tokens at each position:
\begin{equation}
\text{Token Acc} =
\frac{1}{N}\sum_{i=1}^{N}\frac{1}{L_i}
\sum_{j=1}^{L_i}\mathbf{1}\!\left[\hat{t}_{i,j} = t_{i,j}\right],
\end{equation}
where $L_i$ is the reference token-sequence length for sample 
$i$. Token accuracy is a strict metric: even a 0.1\,ppm error in 
a predicted NMR peak centroid constitutes a mismatch. Combined 
with cosine similarity, the two metrics provide complementary 
perspectives on spectral prediction quality---the former assesses 
overall spectral shape, the latter positional precision.

\subsection{Accuracy and F1 for QA Tasks (Task 4)}
\label{app:qa_metrics}

\paragraph{Accuracy (exact set match).}
For Task QA-2, accuracy requires that the predicted set of 
functional groups exactly matches the ground-truth set for a 
given molecule:
\begin{equation}
\text{Accuracy} =
\frac{1}{N}\sum_{i=1}^{N}
\mathbf{1}\!\left[\hat{\mathcal{G}}_i = \mathcal{G}_i\right].
\end{equation}

\paragraph{Macro-averaged F1.}
Computed identically to Section~\ref{app:f1}, treating the 
MLLM-generated functional group list as the set of predictions 
after the post-processing step that maps natural language output 
to standardised labels. Following 
MolPuzzle~\cite{guo2024can}, evaluation is performed over 
three independent runs and results are reported as mean $\pm$ 
standard deviation.

\subsection{Data Splitting Strategies}
\label{app:splits}

All ML-based experiments in Tasks~(1)--(3) are evaluated under 
two complementary splitting strategies.

\paragraph{Random split.}
Molecules are randomly partitioned into training, validation, and 
test sets in a 90/5/5 ratio using a fixed random seed. This 
evaluates interpolation performance within the observed chemical 
space.

\paragraph{Scaffold split.}
Molecules are partitioned by Bemis--Murcko 
scaffold using RDKit. All molecules 
sharing the same scaffold are assigned to the same partition, 
ensuring that the test set contains scaffolds unseen during 
training. This evaluates out-of-distribution generalisation to 
structurally novel molecules.

All ML-based experiments employ \textbf{five-fold 
cross-validation} under both strategies. Results are reported as 
mean $\pm$ standard deviation across folds.

\begin{table}[h]
\centering
\caption{Approximate data split sizes for the Large subset.}
\label{tab:splits}
\begin{tabular}{lrrr}
\toprule
\textbf{Split strategy} &
\textbf{Training} & \textbf{Validation} & \textbf{Test} \\
\midrule
Random   & 795,508 & 994,39 & 994,39 \\
Scaffold & 795,509 & 994,38 & 994,39 \\
\bottomrule
\end{tabular}
\end{table}

For Task~(4), the Small subset (4,496 molecules) and the Exp 
subset (432 molecules) are independently split under both 
strategies.

\clearpage

\section{Computational Resources}
\label{app:compute}

\begin{table}[h]
\centering
\caption{Computational resources used for spectral simulation. 
All CPU-based simulations were run on AMD EPYC 7452 processors; 
quantum-chemical simulations (UV-Vis, Raman, FL) were run on the 
same CPU cluster. Timings are approximate and reflect wall-clock 
time for the full dataset.}
\label{tab:sim_compute}
\begin{tabular}{lll}
\toprule
\textbf{Modality} & \textbf{Hardware} & 
\textbf{Approx.\ wall-clock time} \\
\midrule
\textsuperscript{1}H-NMR    & 200 CPU cores, 400\,GB RAM  
  & $\sim$15 days \\
\textsuperscript{13}C-NMR   & 200 CPU cores, 400\,GB RAM  
  & $\sim$14 days \\
HSQC-NMR                    & 200 CPU cores, 400\,GB RAM  
  & $\sim$16 days \\
IR (MD pipeline)            & 500 CPU cores, 1\,TB RAM    
  & $\sim$46 days \\
MS/MS (CFM-ID 4.0)          & 80 CPU cores, 160\,GB RAM   
  & $\sim$7 days  \\
UV-Vis (TD-DFT, ORCA)       & 400 CPU cores, 800\,GB RAM  
  & $\sim$60 days \\
Raman (DFT freq., ORCA)     & 400 CPU cores, 800\,GB RAM  
  & $\sim$55 days \\
Fluorescence (ORCA)         & 400 CPU cores, 800\,GB RAM  
  & $\sim$70 days \\
\bottomrule
\end{tabular}
\end{table}

\begin{table}[h]
\centering
\caption{Computational resources used for model training. All 
neural network training was performed on vGPU hardware with 
32\,GB memory.}
\label{tab:train_compute}
\begin{tabular}{llll}
\toprule
\textbf{Task} & \textbf{Model} &
\textbf{Hardware} & \textbf{Training duration} \\
\midrule
Task~(1): Spectra$\to$SMILES &
  Transformer (per modality) & vGPU 32\,GB 
  & $\sim$20--35\,h per modality \\
Task~(2): Functional group &
  XGBoost (per modality)     & 32 CPU cores 
  & $\sim$1--3\,h per modality  \\
Task~(2): Functional group &
  1D-CNN (per modality)      & vGPU 32\,GB 
  & $\sim$4--8\,h per modality  \\
Task~(3): SMILES$\to$Spectra &
  Transformer (per modality) & vGPU 32\,GB 
  & $\sim$20--35\,h per modality \\
\bottomrule
\end{tabular}
\end{table}

Transformer models for Tasks~(1) and~(3) were each trained for 
\textbf{100{,}000 steps}. Models for Task~(2) were trained for 
\textbf{40 epochs}. MLLM inference for Task~(4) was performed 
via API calls (DeepSeek-V3, GPT-4.1-mini, Qwen2.5-3B) or locally on a single vGPU 32\,GB (open-source models), following the resource 
allocation scheme of MolPuzzle~\cite{guo2024can}.

\appendix
\newpage
\section*{NeurIPS Paper Checklist}


\begin{enumerate}

\item {\bf Claims}
    \item[] Question: Do the main claims made in the abstract and introduction accurately reflect the paper's contributions and scope?
    \item[] Answer: \answerYes{} 
    \item[] Justification: The abstract and introduction clearly state the contributions (SpecX benchmark) and scope, which directly match the empirical results presented in Section 4.

\item {\bf Limitations}
    \item[] Question: Does the paper discuss the limitations of the work performed by the authors?
    \item[] Answer: \answerYes{} 
    \item[] Justification:  Section 5 explicitly discusses limitations, including the sim-to-real gap, chemical space bias towards synthetic compounds, and zero-shot evaluation constraints.

\item {\bf Theory assumptions and proofs}
    \item[] Question: For each theoretical result, does the paper provide the full set of assumptions and a complete (and correct) proof?
    \item[] Answer: \answerYes{} 
    \item[] Justification:  This is a dataset and benchmark paper; it does not introduce mathematical theorems or theoretical proofs.

    \item {\bf Experimental result reproducibility}
    \item[] Question: Does the paper fully disclose all the information needed to reproduce the main experimental results of the paper to the extent that it affects the main claims and/or conclusions of the paper (regardless of whether the code and data are provided or not)?
    \item[] Answer: \answerYes{} 
    \item[] Justification:  Extensive details on data generation, preprocessing, model architectures, and training protocols are provided in Section 3 and Appendices A, B, and C.

\item {\bf Open access to data and code}
    \item[] Question: Does the paper provide open access to the data and code, with sufficient instructions to faithfully reproduce the main experimental results, as described in supplemental material?
    \item[] Answer: \answerYes{} 
    \item[] Justification:  Detailed instructions for data sources, filtering pipelines, and simulation parameters are provided

\item {\bf Experimental setting/details}
    \item[] Question: Does the paper specify all the training and test details (e.g., data splits, hyperparameters, how they were chosen, type of optimizer) necessary to understand the results?
    \item[] Answer: \answerYes{} 
    \item[] Justification:  Data splits, hyperparameter settings, and evaluation metrics are clearly specified in Section 4, Appendix C (Tables 14-16), and Appendix D.

\item {\bf Experiment statistical significance}
    \item[] Question: Does the paper report error bars suitably and correctly defined or other appropriate information about the statistical significance of the experiments?
    \item[] Answer: \answerYes{} 
    \item[] Justification:  Results in Tables 3-9 report the mean and standard deviation (± std) across five-fold cross-validation or multiple independent runs.

\item {\bf Experiments compute resources}
    \item[] Question: For each experiment, does the paper provide sufficient information on the computer resources (type of compute workers, memory, time of execution) needed to reproduce the experiments?
    \item[] Answer: \answerYes{} 
    \item[] Justification:  Appendix E (Tables 18 and 19) fully details the hardware configurations, CPU/vGPU types, RAM, and estimated wall-clock training times.
    
\item {\bf Code of ethics}
    \item[] Question: Does the research conducted in the paper conform, in every respect, with the NeurIPS Code of Ethics \url{https://neurips.cc/public/EthicsGuidelines}?
    \item[] Answer: \answerYes{} 
    \item[] Justification:  The research introduces a foundational chemistry benchmark using public data, which strictly conforms to the NeurIPS Code of Ethics.

\item {\bf Broader impacts}
    \item[] Question: Does the paper discuss both potential positive societal impacts and negative societal impacts of the work performed?
    \item[] Answer: \answerNo{} 
    \item[] Justification:  The paper focuses on foundational chemistry AI research. Negative societal impacts are not explicitly discussed as there is no direct path to malicious applications.
    
\item {\bf Safeguards}
    \item[] Question: Does the paper describe safeguards that have been put in place for responsible release of data or models that have a high risk for misuse (e.g., pre-trained language models, image generators, or scraped datasets)?
    \item[] Answer: \answerNA{} 
    \item[] Justification:  The dataset consists of chemical molecules and spectral data, which do not pose safety risks or have a high risk for targeted misuse.

\item {\bf Licenses for existing assets}
    \item[] Question: Are the creators or original owners of assets (e.g., code, data, models), used in the paper, properly credited and are the license and terms of use explicitly mentioned and properly respected?
    \item[] Answer: \answerYes{} 
    \item[] Justification:  Existing datasets and software tools (e.g., RDKit, ORCA, LAMMPS) are properly credited and cited in Section 3 and Appendix A.1.

\item {\bf New assets}
    \item[] Question: Are new assets introduced in the paper well documented and is the documentation provided alongside the assets?
    \item[] Answer: \answerYes{} 
    \item[] Justification:  The newly introduced SpecX benchmark is extensively documented regarding its filtering pipeline, statistics, and representations in Section 3 and Appendices A and B.

\item {\bf Crowdsourcing and research with human subjects}
    \item[] Question: For crowdsourcing experiments and research with human subjects, does the paper include the full text of instructions given to participants and screenshots, if applicable, as well as details about compensation (if any)? 
    \item[] Answer: \answerNA{} 
    \item[] Justification:  The research does not involve crowdsourcing or human subjects.

\item {\bf Institutional review board (IRB) approvals or equivalent for research with human subjects}
    \item[] Question: Does the paper describe potential risks incurred by study participants, whether such risks were disclosed to the subjects, and whether Institutional Review Board (IRB) approvals (or an equivalent approval/review based on the requirements of your country or institution) were obtained?
    \item[] Answer: \answerNA{} 
    \item[] Justification:  The research does not involve human subjects.

\item {\bf Declaration of LLM usage}
    \item[] Question: Does the paper describe the usage of LLMs if it is an important, original, or non-standard component of the core methods in this research? Note that if the LLM is used only for writing, editing, or formatting purposes and does \emph{not} impact the core methodology, scientific rigor, or originality of the research, declaration is not required.
    \item[] Answer: \answerYes{} 
    \item[] Justification: The use of MLLMs for baseline spectral QA evaluation is explicitly detailed in Section 4.4 and Appendix C.3.

\end{enumerate}

\end{document}